# Long-term performance of non-ALD MCP-PMTs in the high-radiation environment of ALICE

Yury Melikyan for the ALICE Collaboration

*Helsinki Institute of Physics, University of Helsinki, P.O. Box 64, FI-00014, Finland*

## ABSTRACT

The Fast Interaction Trigger detector of the ALICE apparatus at the LHC includes a Cherenkov subsystem utilizing 52 Planacon XP85002/FIT-Q microchannel plate-based photomultiplier tubes. It is a precise trigger, multiplicity and luminosity detector located in the forward region, where the flux of collision products is the largest. The photon load reaches up to $2{\cdot}10^8$ photoelectrons / $cm^2$ / s for the innermost photosensors, corresponding to 0.4 μA/$cm^2$ average anode current density. The lifetime-extended version of the Planacon MCP-PMT with atomic layer deposition cannot handle such a large anode current. We therefore use the non-ALD version of the Planacons that, on the other hand, suffer from a limited lifetime in terms of the integrated anode charge.

Here we present the long-term performance trends of the non-ALD Planacons up to the integrated anode charge of 2 C/$cm^2$. We demonstrate that the low-gain operation strategy allows to maintain excellent time resolution in multiphoton mode unchanged up to that large integrated charge. While the majority of the equally illuminated MCP-PMTs feature almost identical ageing speed, two outliers suffering from a faster ageing coincide with the outliers in the noise characteristics measured prior to the irradiation. Moreover, we report that the effective ageing speed is partially suppressed by the self-recovery of the response of aged MCP-PMTs newly observed by our group during months-long no-beam periods.

## 1. Introduction.

The Fast Interaction Trigger (FIT) detector of the ALICE apparatus at the LHC includes a Cherenkov subsystem (FT0) composed of two arrays of 24 (FT0-A) and 28 (FT0-C) independent Cherenkov modules (see Fig.1-2). The two arrays are asymmetrically located at both sides of the collision point of the LHC beams to monitor their collision time, multiplicity, vertex position, and luminosity [1].

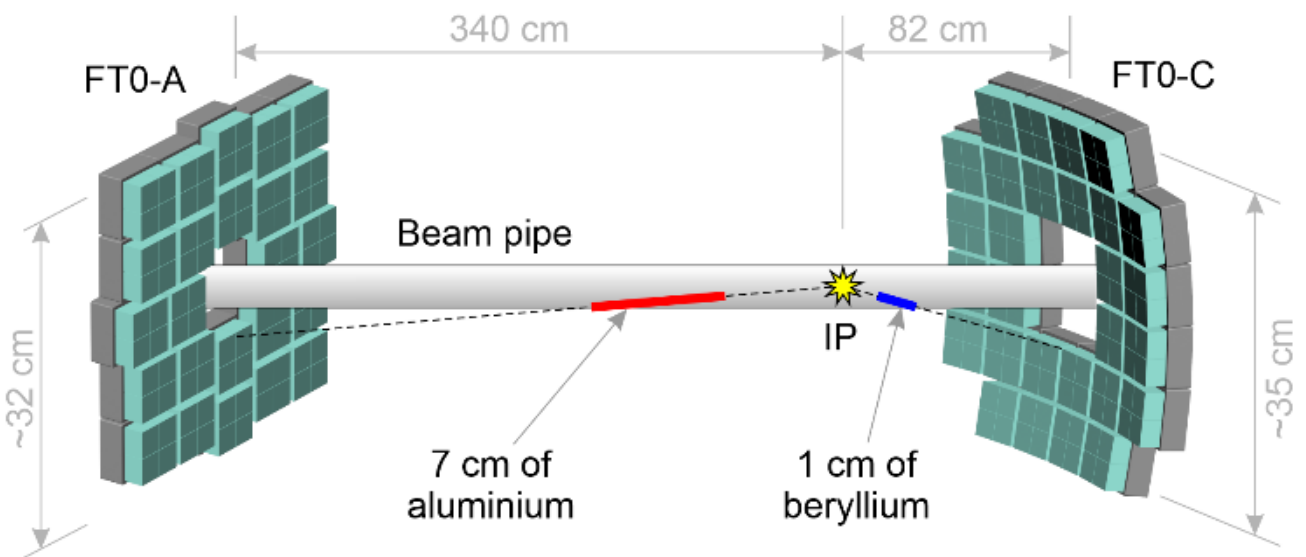

Figure 1. Schematic of the two FIT-FT0 Cherenkov arrays (drawn not to scale). The material budget in front of the innermost FT0-A channels includes 7 cm of the aluminium beampipe and the FV0 scintillation detector (not shown), in front of FT0-C – 1 cm of the beryllium beampipe and the MFT tracker (not shown) [1, 2].

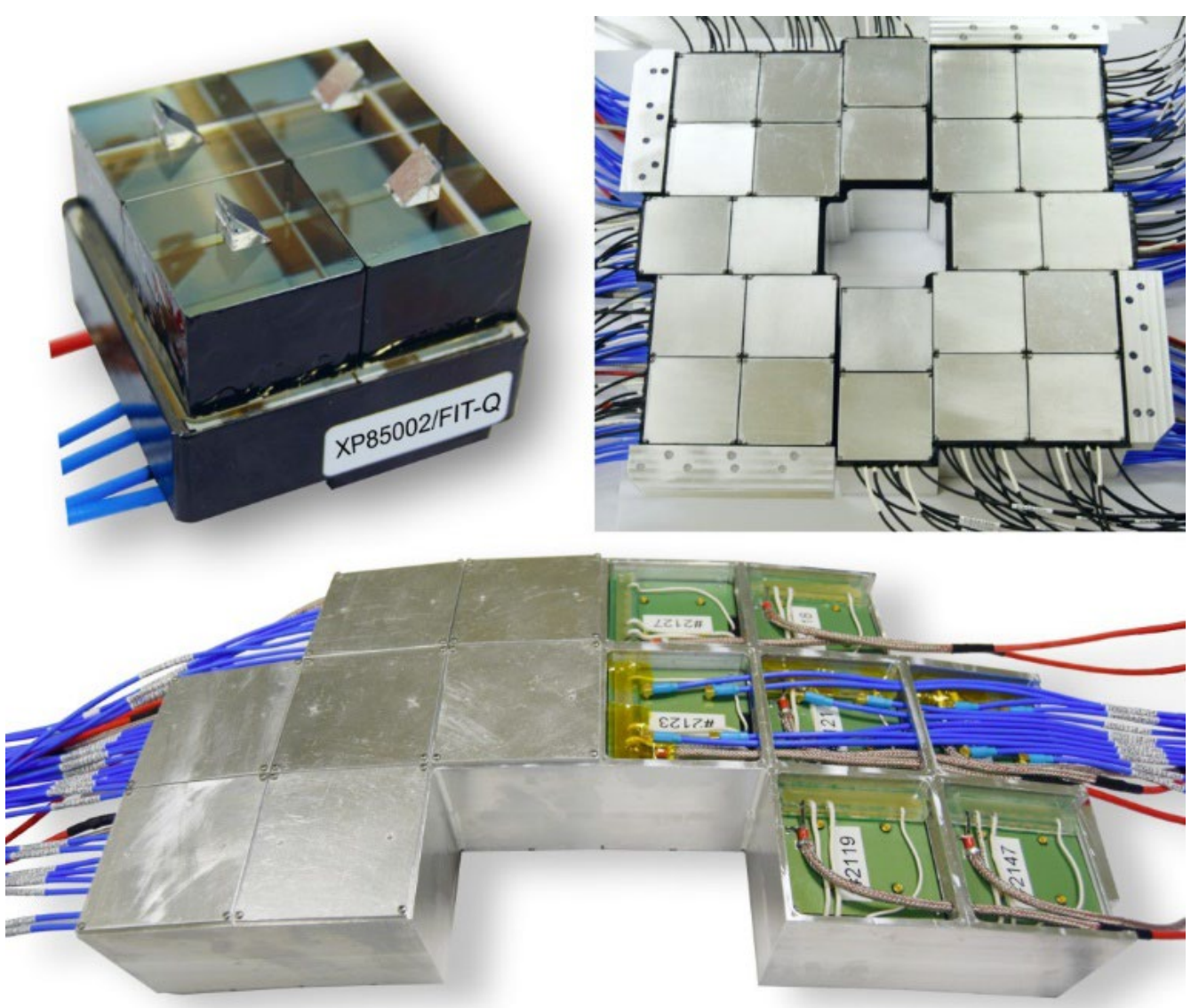

Figure 2. Top-left: photo of one FT0 Cherenkov module; top-right: front view of FT0-A fully assembled; bottom: back view of FT0-C semi-assembled.

Each FT0 Cherenkov module is based on four square-shaped quartz radiators made of fused silica, optically coupled to a multianode microchannel plate-based photomultiplier tube (MCP-PMT). The radiators are 20 mm-thick, with their four side edges mirror-coated. The MCP-PMTs are customized to detect relativistic charged particles with excellent time resolution ($\sigma$ = 13 ps) and wide rate capability – up to $2{\cdot}10^8$ photoelectrons per second, or 0.4 $\mu$A/cm$^2$ average anode current (AAC) density [3]. For the Planacon MCP-PMTs used, this is only achievable without using the Atomic Layer Deposition (ALD) technique intended to increase the device's lifetime, but decreasing the effective average anode current limit as a side effect [4].

## 2. MCP-PMT saturation by AAC

FIT Planacons use chevron stacks of a pair of 1 mm-thick MCPs with 25 $\mu$m pores arranged with 32 $\mu$m pitch. We use low-resistance MCP stacks to increase the chances of having the sufficiently high AAC saturation limit – see Fig.3. The AAC saturation manifests itself at increased photon loads while operating at a given gain – as defined by the imbalance between the anode current and MCP strip current recharging the pores [5].

Most part of the running time, the ALICE solenoid enclosing both FT0 detectors provides a magnetic field of B = 0.5 T, while running at B = 0.2 T happens occasionally. Figure 4 shows an example of the AAC saturation curves as measured for one MCP-PMT with $R_{MCP}$ = 15 M$\Omega$ at various magnetic fields. Presence of magnetic field increases the number of interactions of the secondary electrons with the MCP pore walls, affecting the secondary electron yield [6] and consequently the MCP gain. MCP-PMT bias voltage was therefore adjusted for each B-field setting to conduct the AAC saturation measurement at the same electron gain ($1.5{\cdot}10^4$) and light intensity of the input pulse.

As can be seen from Fig.4, both at B = 0.2 T and B = 0.5 T the overlinearity effect [7] diminishes, and the AAC limit decreases. While the exact mechanism requires further study, it may be related to a magnetic-field-induced redistribution of the avalanche development inside the MCP pores. There is no significant effect on the time resolution throughout the entire AAC range depicted in Fig.4, extra time delay in excess of ~100 ps is observed to evolve rapidly once the MCP-PMT gain drops to 1/3 of the initial value.

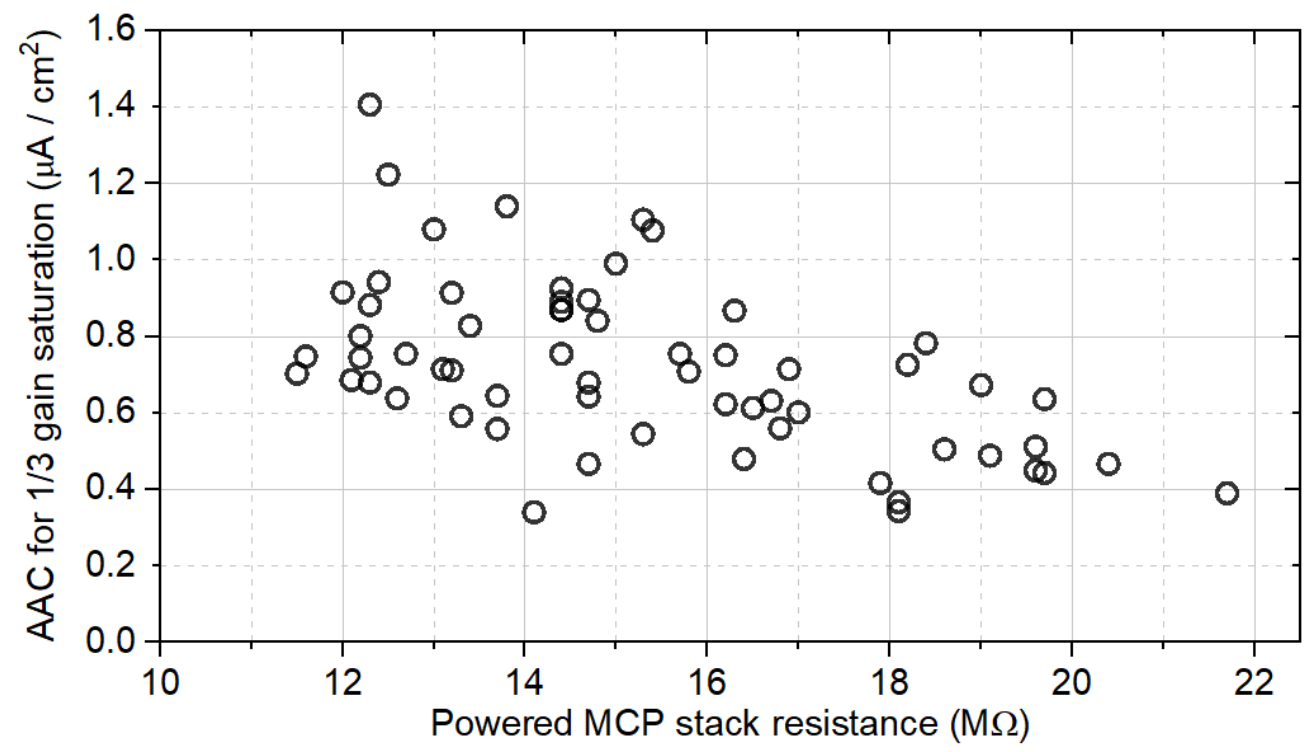

Figure 3. Dependence of the AAC limit for the MCP-PMT gain saturation by 1/3 versus the MCP stack resistance of the Planacons produced for ALICE FIT. Same data in dependence of the MCP strip current is provided in [7].

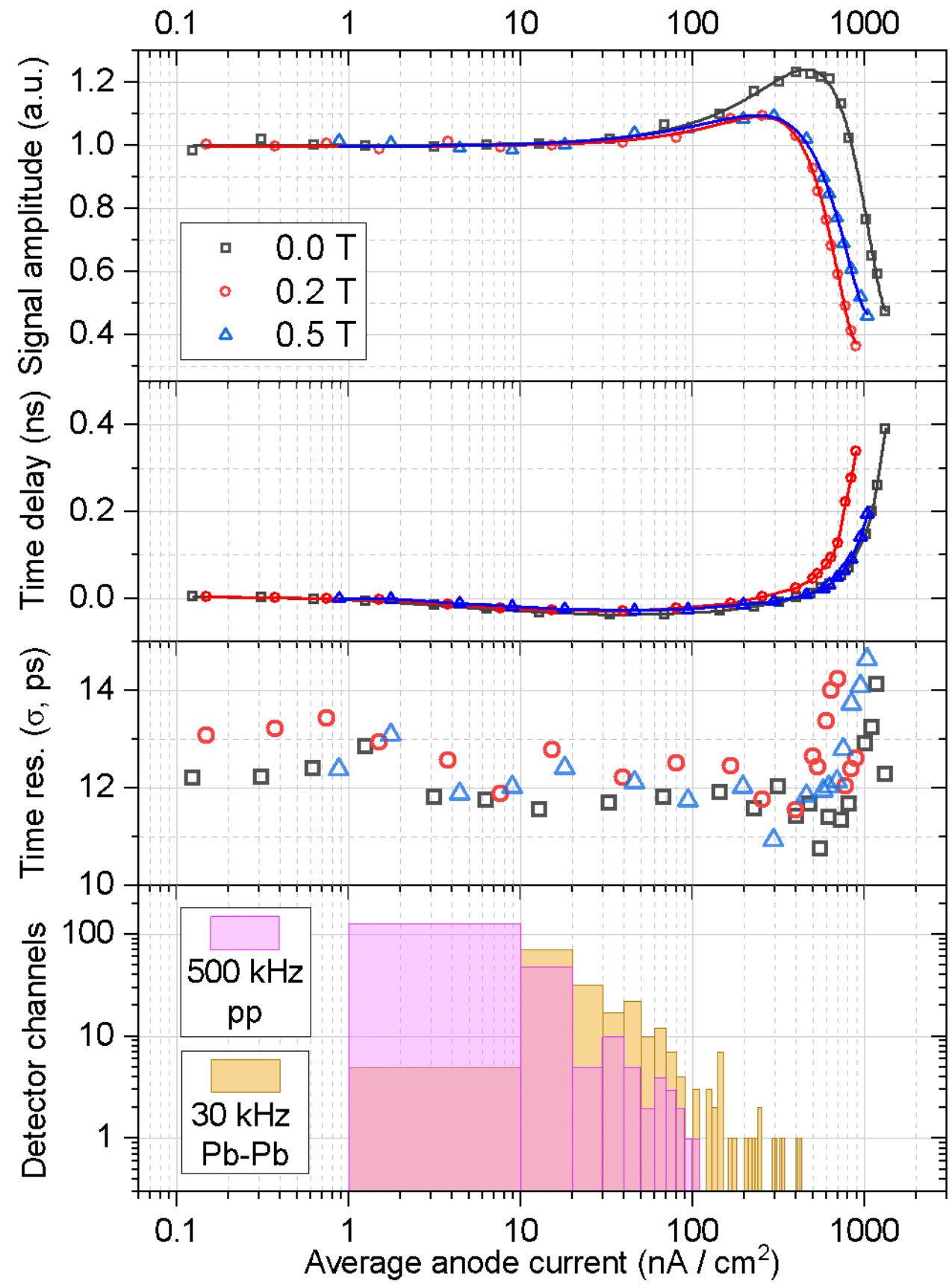

Figure 4. Saturation curves of a sample MCP-PMT as a function of the AAC density at different magnetic fields (perpendicular to the MCP plane). MCP-PMT bias voltage was adjusted individually for each B-field to maintain the same electron gain at low AAC. Signal amplitude of each saturation curve is normalized to its mean value in the range of AAC < 5 nA/cm$^2$. AAC was altered by varying the repetition rate of light pulses of unchanged intensity in the range from ~20 Hz to ~400 kHz. The bottom section shows the real distribution of the FT0 per-channel AAC values as seen at 500 kHz pp and 30 kHz Pb-Pb.

Trigger threshold of each FT0 channel is set to ~60% of single-MIP (Minimum Ionizing Particle) energy deposition. With ~70% single-MIP energy resolution (calculated as FWHM / mean), ~7% of single-MIP distribution lags behind the trigger threshold even at low AAC. Proton-proton collision flag requires a coincidence between triggers from FT0-A and FT0-C. On average, one proton-proton collision results in ~50 particles detected in FT0-A and ~10 particles detected in FT0-C. For the proton-proton colliding system where FT0 serves as the primary luminometer of ALICE, it reduces the number of undetected collisions to negligible values.

The maximum collision rate of Pb-Pb beams at ALICE did not exceed 30 kHz in 2025. These conditions cause ~4 times higher photon load to the MCP-PMTs than the 500 kHz collision rate of proton beams.

Even under the 30 kHz Pb-Pb beams, we avoid any significant saturation of FT0. As can be seen from Fig.5, $2{\cdot}10^{8}$ p.e. / cm$^2$ / s photon load in the innermost FT0-A MCP-PMTs causes 2…12% of extra loss in per-channel trigger efficiency corresponding to 7…23% drop in gain. Other channels stay unaffected – the number of secondaries seen by the innermost FT0-A channels is inflated because of the larger material budget as defined by their pseudorapidity (see Fig.1). In Pb-Pb colliding system where FT0 serves as the secondary luminometer, the large number of particles – collision products reduces the total efficiency loss to even smaller values.

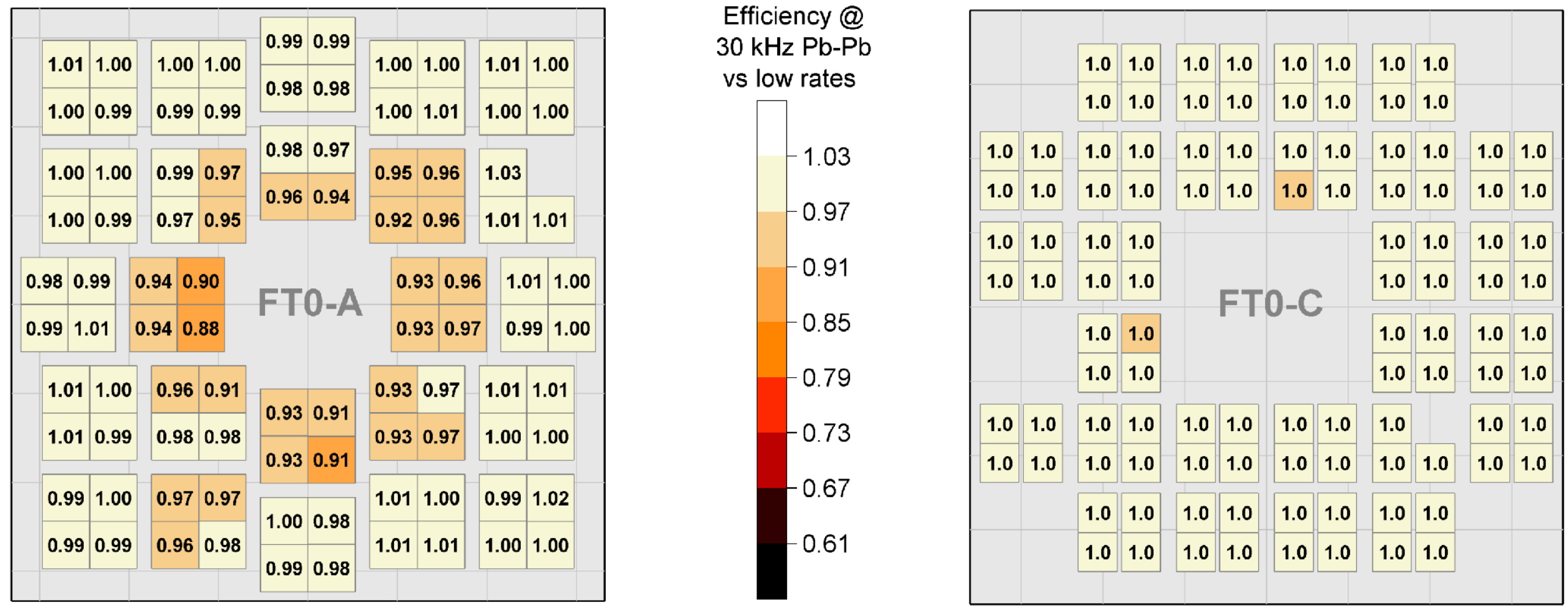

Figure 5. Effect of the AAC saturation to the per-channel trigger efficiency as seen under the most intense load coming from a 30 kHz Pb-Pb collision rate.

**3. MCP-PMT ageing**

Since the detector deployment back in 2021, innermost FT0 MCP-PMTs have been exposed to ~$10^{12}$ 1-MeV $n_{eqv}$/cm$^2$ hadron fluence and up to ~100 kRad of total ionizing dose [8]. The former value is simulated for a sample silicon chunk replacing FT0, according to the non-ionizing energy loss (NIEL) hypothesis and hardness factors for silicon [9]. However, large contribution to the FT0-C comes from backflow of secondaries produced by the massive muon absorber right behind [2]. As the front surface of the FT0 quartz radiators (one edge) is put into optical contact with a light absorbing layer, backward Cherenkov light does not contribute to the integral anode charge (IAC), keeping it ~5 times lower for FT0-C than for FT0-A.

Planacon MCP-PMTs are radiation-hard detectors, but their lifetime is limited in terms of IAC. Up to 2 C/cm$^2$ has been accumulated at the innermost FT0-A MCP-PMTs so far. Change in the response has been continuously monitored for all MCP-PMTs throughout the detector's lifetime with the help of a dedicated radiation-hard optical system equipped with a blue pulsed laser and a reference Planacon MCP-PMT. The latter is placed outside the irradiated area and is not equipped with Cherenkov radiators. It is capable to correct for the deviations in the laser light intensity, and to monitor for possible changes in the transparency of the optical system path.

Change in the FT0 MCP-PMT response by September 2025 is mapped in Fig.6. It depicts the change in the output charge for a given number of injected photons, thus reflecting the change of the quantum efficiency convoluted with gain. Even though the response of the most loaded units was suppressed down to ~40% of the original, precise timing performance of the Cherenkov detector remains unaffected so far. A time resolution of $\sigma$ = ~15 ps for ~10 MIP-signals allows FT0 to resolve proton collisions with $\sigma$ = 17 ps time resolution, Pb-Pb ion collisions – with $\sigma$ = 4.4 ps resolution [1]. This temporal characteristic, as well as the detector efficiency [1], stay unchanged in spite of the significant MCP-PMT ageing – thanks to the regular increase in the MCP-PMT bias voltage. Its typical value for the most loaded channels grew from ~1.3 kV to ~1.4 kV over the entire period of detector operation. Ageing of the MCP-PMT response is monitored with laser outside the beam time. MCP-PMT bias voltage settings are kept unchanged during the laser scans for a proper ageing monitoring. Ageing of the inner quadrants of each FT0-A MCP-PMT is shown in Fig.7 as a function of IAC.

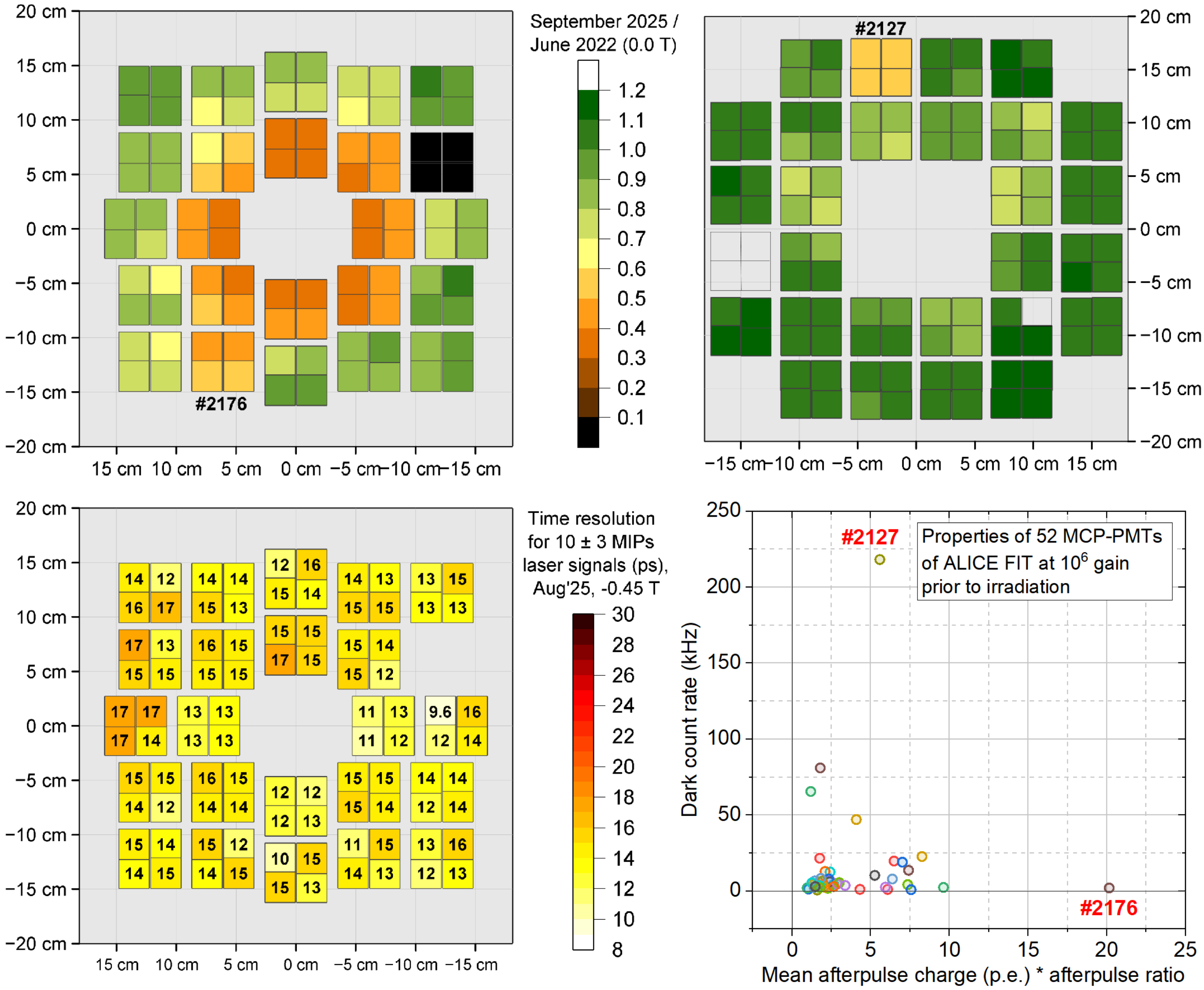

Figure 6. Top: change in the MCP-PMT response to blue light with the bias voltage unchanged since the first stable beams of RUN3 LHC. The only blind FT0-A unit suffered from a vacuum microleak, and a missing FT0-C unit suffered from a non-recoverable HV breakdown across the MCP. Bottom-left: time resolution of “aged” FT0-A channels, showing no correlation to actual ageing once the response is compensated with the elevated bias voltage. Note FT0 is optimized to detect signals in the range 1-260 MIPs. Bottom-right: noise characteristics of the FIT Planacons prior to their irradiation [7]. The two outliers in noise level coincide with the two outliers in the ageing speed.

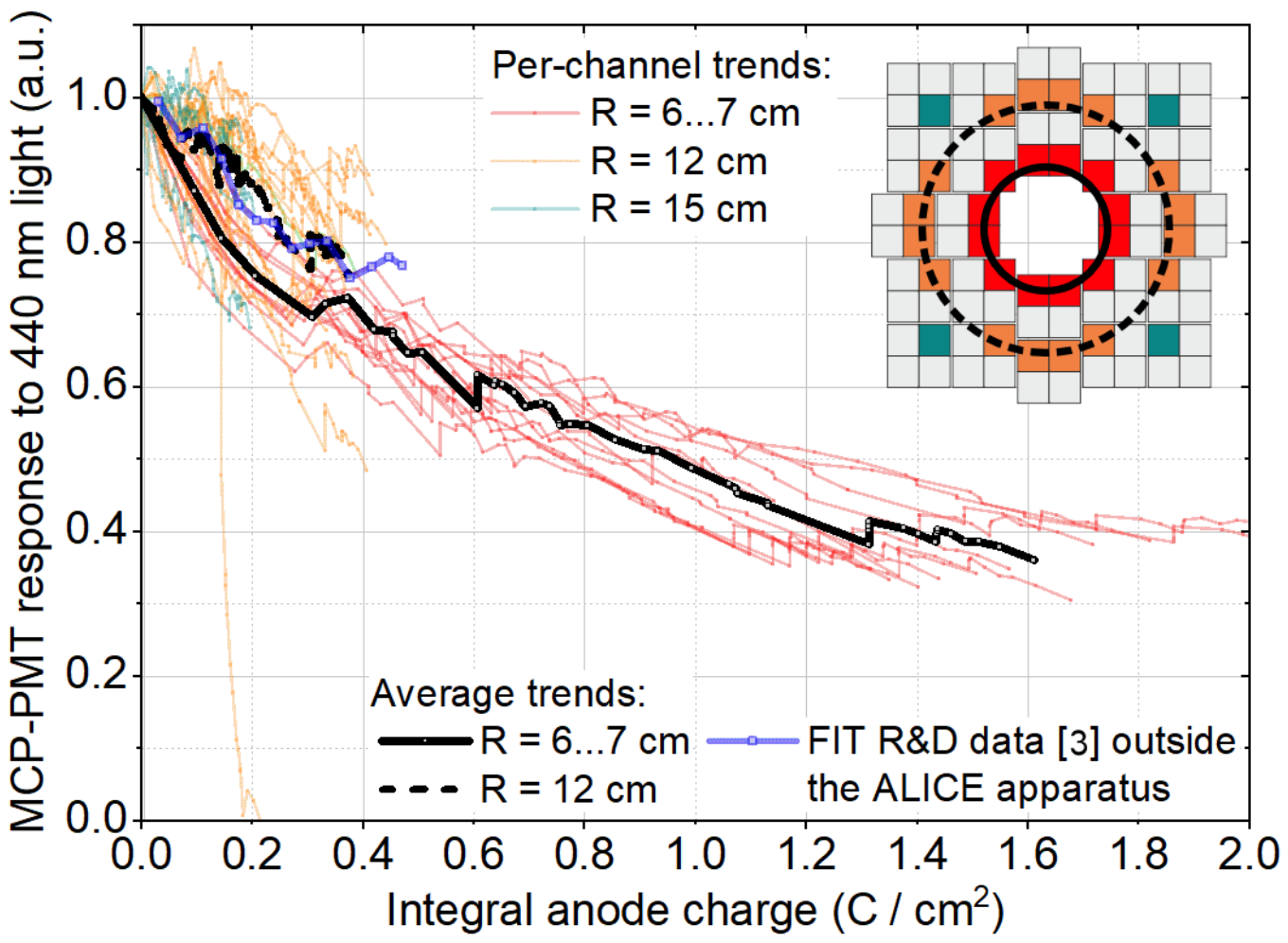

Figure 7. Change of the response of the inner quadrants of each FT0-A MCP-PMT as a function of IAC. Bias voltage was kept unchanged within the plotted interval. The two black curves show average trends

among the equally-illuminated quadrants (located at a similar radius from the beam axis). The average ageing trend of the moderately-illuminated quadrants overlaps well with the ageing trend measured under pure LED light outside the ALICE apparatus [3].

### 4. Self-recovery of aged MCP-PMTs

The ageing shown in Fig.7 features a clear downtrend, but a few rapid positive corrections can be noticed. Their horizontal position coincides with the IAC values accumulated by the start of the prolonged no-beam periods typical for the year-end shutdowns of the LHC. Starting from the second year of RUN3, we performed regular monitoring of the MCP-PMT response also during the no-beam periods. It allowed us to observe steady uptrends in the response of all aged MCP-PMTs in absence of the regular illumination by Cherenkov light – see Fig.8-9.

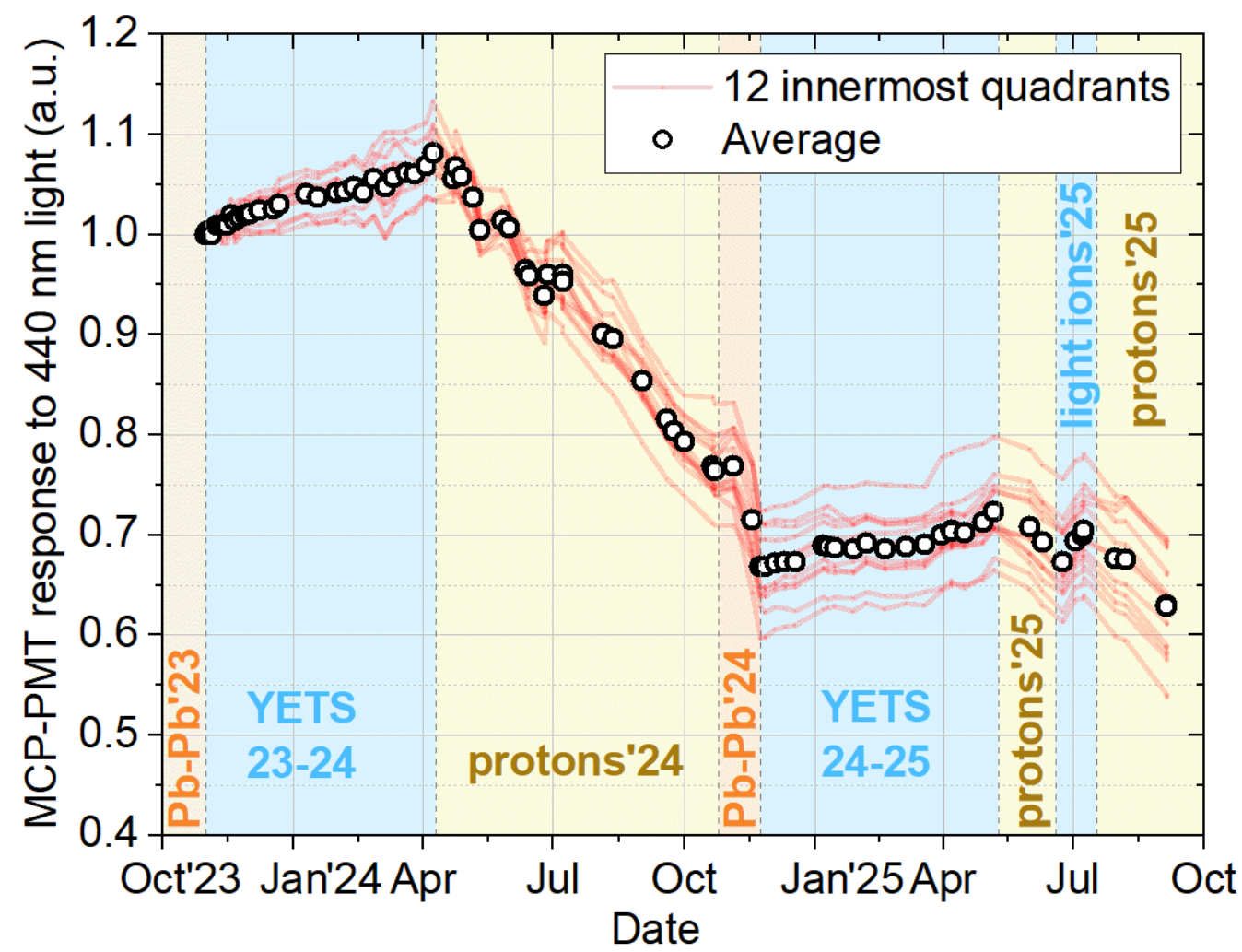

Figure 8. Change of the MCP-PMT response as a function of time at bias voltage unchanged within the plotted interval. Periods of active MCP-PMT illumination with Cherenkov light are highlighted with yellow and orange. Blue areas highlight the year-end technical stops (YETS) of the LHC and the low-lumi periods of inactive illumination. All data points were measured at B = 0.0 T.

During the two 160 days-long no-beam periods, the response of aged MCP-PMTs self-recovered by ~8%, while the response of those MCP-PMTs not affected by ageing did not change significantly. Temperature of the FT0 environment is permanently kept at 20 ± 1 (°C) indpendently of the LHC operation mode. The monitoring light is injected to all $(24 + 28) \cdot 4 = 208$ FT0 channels by a single laser, being delivered to each individual MCP-PMT quadrant by a system of rad-hard fibers. The fiber transparency is stable within ±1% during each of the YETS periods. Being a key ALICE detector, FT0 cannot be subject to invasive tests until the end of RUN3 in mid-2026 – until then, exact nature of the self-recovery effect remains unclear.

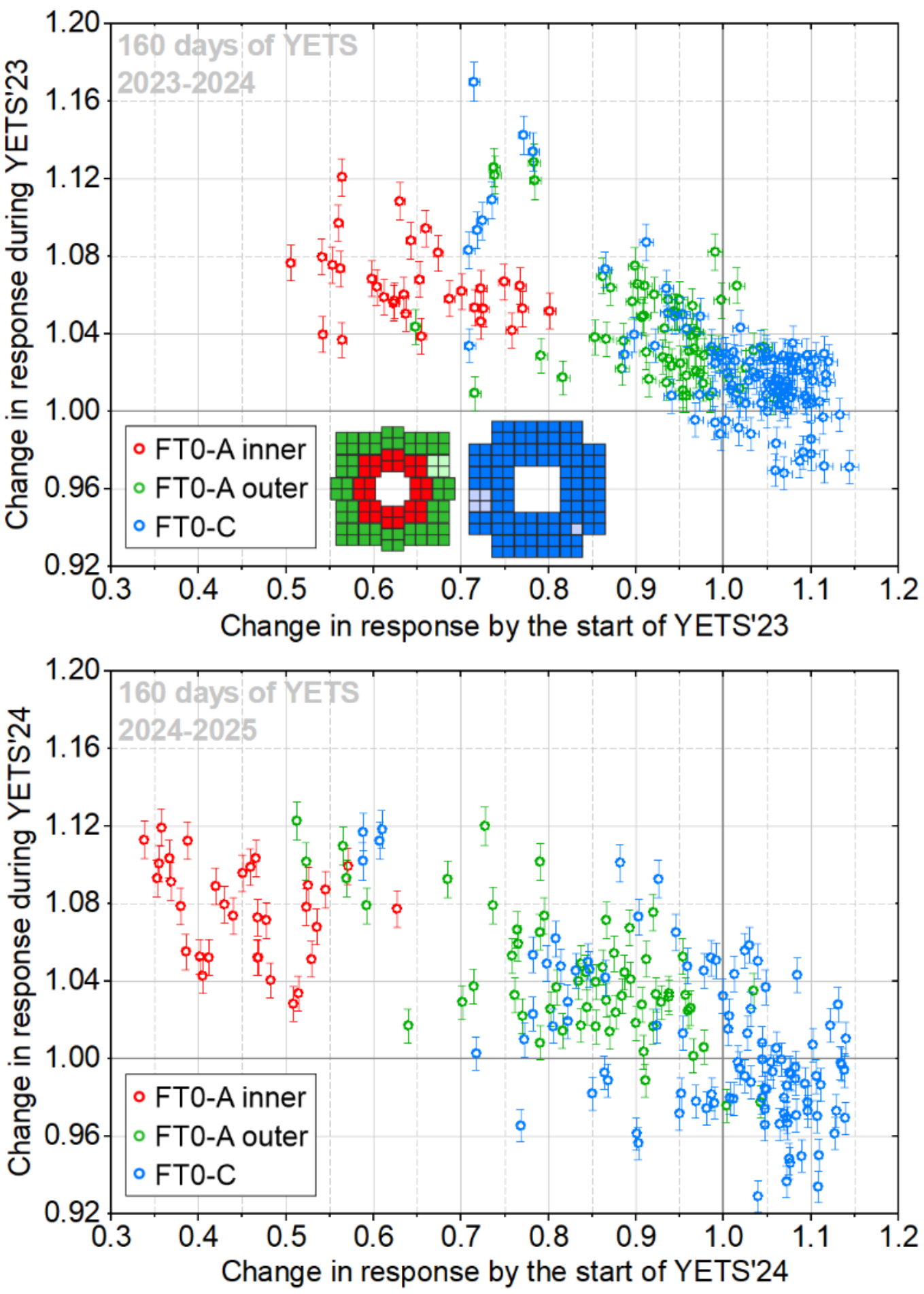

Figure 9. Value of the self-recovery effect as a function of the preceding ageing effect – data from the two plots were collected one year apart from each other. Those MCP-PMTs not affected by ageing do not show any increase in the response during YETS, while aged MCP-PMTs do.

## 5. Conclusions

ALICE FIT serves as the first large-scale use case for the Planacon MCP-PMTs in high-energy physics. The choice of the non-ALD MCP-PMT option allowed us to operate under AAC of ~0.4 μA/cm$^2$ in a magnetic field of B = 0.5 T with no significant signs of saturation. In spite of the limited lifetime in terms of IAC, these devices can be successfully used in the forward environment of ALICE for high-speed counting of relativistic particles provided that a sufficient margin in Cherenkov yield is secured by the chosen radiator type and size. In our case, 20 mm-thick quartz radiators allow us to maintain remarkably high Pb-Pb collision time resolution of σ = 4.4 ps even after the response of the most loaded photosensors is affected by ageing.

A self-recovery effect of aged MCP-PMTs was observed during the extended no-beam periods – its nature will be clarified once the most aged MCP-PMTs are extracted from the ALICE FIT detector upon the end of the LHC RUN3.